\documentclass{webofc}

\usepackage[varg]{txfonts}   
\usepackage{graphicx}
\usepackage{hyperref}
\usepackage{url}
\hypersetup{colorlinks=true,citecolor=blue,urlcolor=blue,linkcolor=blue}

\begin{document}

\title{Status and Prospects of the HEP Statistical Inference Ecosystem}

\author{\firstname{Massimiliano} \lastname{Galli}\inst{1}\fnsep\thanks{\email{mg2708@princeton.edu}}
}

\institute{Princeton University, Princeton, NJ, USA}

\abstract{%
Statistical inference is a crucial part of HEP analyses. Historically based on
RooFit and RooStats, the statistical tools used by the experiments are now facing
unprecedented challenges, such as the rapidly growing complexity of statistical
models (involving hundreds of parameters of interest and thousands of nuisance
parameters), the need for scalable performance in large likelihood minimizations,
and the demand for interoperability across an increasingly diverse ecosystem of
tools, computational hardware, and frameworks and libraries, especially the ones
developed and used within the machine learning world.
This contribution summarizes status and future plans for the statistical tools used
by some of the main LHC experiments (CMS, ATLAS), with a focus on improvements
coming from the ROOT world (RooFit automatic differentiation), interoperability with
modern libraries (JAX) and communication across frameworks (HS3).
}

\maketitle

\section{Introduction}
\label{sec:intro}

Statistical inference is a key step in deriving results in high energy physics.
Data and a statistical model are combined into a likelihood function, from which
limit setting, parameter estimation and discovery claims all follow through
minimization and hypothesis testing. The statistical model encodes how expected
observations depend on the parameters of interest and on the nuisance parameters
describing systematic uncertainties, and it therefore carries the physics content
of an analysis in a form that can be reused, combined and reinterpreted long after
the original measurement~\cite{Cranmer:2021urp}.

These models are growing quickly. The CMS model used for the discovery of the Higgs
boson contained of order 800 parameters~\cite{cms_collaboration_2024_c2948-e8875}; the recent CMS combination of Higgs boson measurements over the full Run~2 dataset involves more than 1000 histograms, in
excess of 10000 parameters and more than 90 parameters of interest, spanning
template and parametric models across many decay channels~\cite{CMS:2025jwz}.
The High-Luminosity LHC (HL-LHC) will push this complexity further still, through larger datasets, a richer treatment of systematic uncertainties, more sophisticated fits, and combinations across different runs and experiments.

Four pressing needs follow from this trend: \emph{scalable performance}, so that
fits of realistic models remain tractable; \emph{publication and preservation} of
the statistical models themselves, so that results can be reused; \emph{interoperability}
between the frameworks in which those models are built; and \emph{integration with
machine learning}, as techniques such as simulation-based inference enter mainstream
analyses. These four themes were the focus of the IRIS-HEP~\cite{IRISHEPSPUPDATE}
Statistical Ecosystem Blueprint workshop, held at CERN on 24--25 February 2026,
which brought together developers of both the ROOT-based~\cite{Brun:1997pa} and the
Python-based statistical tooling ecosystems. The present contribution is largely
based on the status and the R\&D directions identified there.

This conference paper is organized as follows. Section~\ref{sec:today} surveys the LHC
statistical ecosystem as it stands. Section~\ref{sec:ad} describes automatic
differentiation in RooFit. Section~\ref{sec:hs3} covers model preservation and
the HEP Statistics Serialization Standard. Section~\ref{sec:experiments} presents
the setups and roadmaps of CMS and ATLAS. Section~\ref{sec:jax} discusses the
emerging JAX-based ecosystem and its interoperability with RooFit, and
Section~\ref{sec:sbi} the integration with simulation-based inference workflows.
Section~\ref{sec:summary} summarizes.

\section{The LHC statistical ecosystem today}
\label{sec:today}

The statistical tooling used at the LHC falls broadly into two groups.

The first is built on ROOT. RooFit~\cite{Verkerke:2003ir} and
RooStats~\cite{Moneta:2010pm} provide the modelling and inference layer, with
Minuit2~\cite{Hatlo:2004sga} performing the minimization. On top of this sit the
experiment-facing tools: CMS Combine~\cite{CMS:2024onh} and, in ATLAS,
\texttt{TRExFitter}, \texttt{HistFitter} and others. Taken together, this stack is
cited by several hundred analyses per year and remains the backbone of LHC
statistical analysis.

The second group is the Python ecosystem: \texttt{pyhf}~\cite{pyhf_joss},
\texttt{cabinetry}, \texttt{zfit}~\cite{Eschle:2019jmu},
\texttt{evermore}~\cite{evermore} and others. These are used by of order tens of
analyses per year, a smaller but steadily growing share, and they are the venue in
which much of the recent methodological experimentation has taken place.

The rest of this contribution follows from this split, examining how these tools
are adapting to the needs listed in Section~\ref{sec:intro}, what is new in the
field, and whether the experiments are keeping up with those developments.

\section{Scalable performance: automatic differentiation in RooFit}
\label{sec:ad}

Three main bottlenecks limit minimization performance in the RooFit/Minuit2 stack:
likelihood evaluation, gradient computation, and the linear algebra internal to
Minuit2. Of these, gradient computation dominates for models with many parameters.
Minuit2 computes gradients numerically by default, varying each parameter
independently, so that the cost of a single gradient scales linearly with the
number of parameters. RooFit mitigates this in part through caching of intermediate
results, but for models with hundreds or thousands of parameters, a regime
already common in Run~3, the gradient remains the leading cost, accounting for
up to roughly 90\% of the total fitting time.

Reverse-mode automatic differentiation (AD) addresses this scaling directly: the
cost of evaluating the full gradient differs from the cost of evaluating the
function itself only by a constant factor, independent of the number of parameters.
RooFit integrates an AD engine based on Clad~\cite{Singh:2023xgx}, a source-code
transformation tool implemented as a Clang compiler plugin.

The integration relies on a code generation (\textit{codegen}) pipeline. RooFit
first translates its computation graph into a standalone C\texttt{++} function
that encodes only the mathematical operations of the likelihood. This intermediate
product is valuable in its own right: the generated code carries no dependence on
RooFit, which makes the model both more transparent and more straightforwardly
preservable. Clad then differentiates this code to produce an analytic gradient,
which RooFit wraps back into an object usable by Minuit2. From the user's point of
view the whole mechanism is activated by a single extra flag in the fit call.

\begin{figure}[htbp]
\centering
\includegraphics[width=0.48\textwidth]{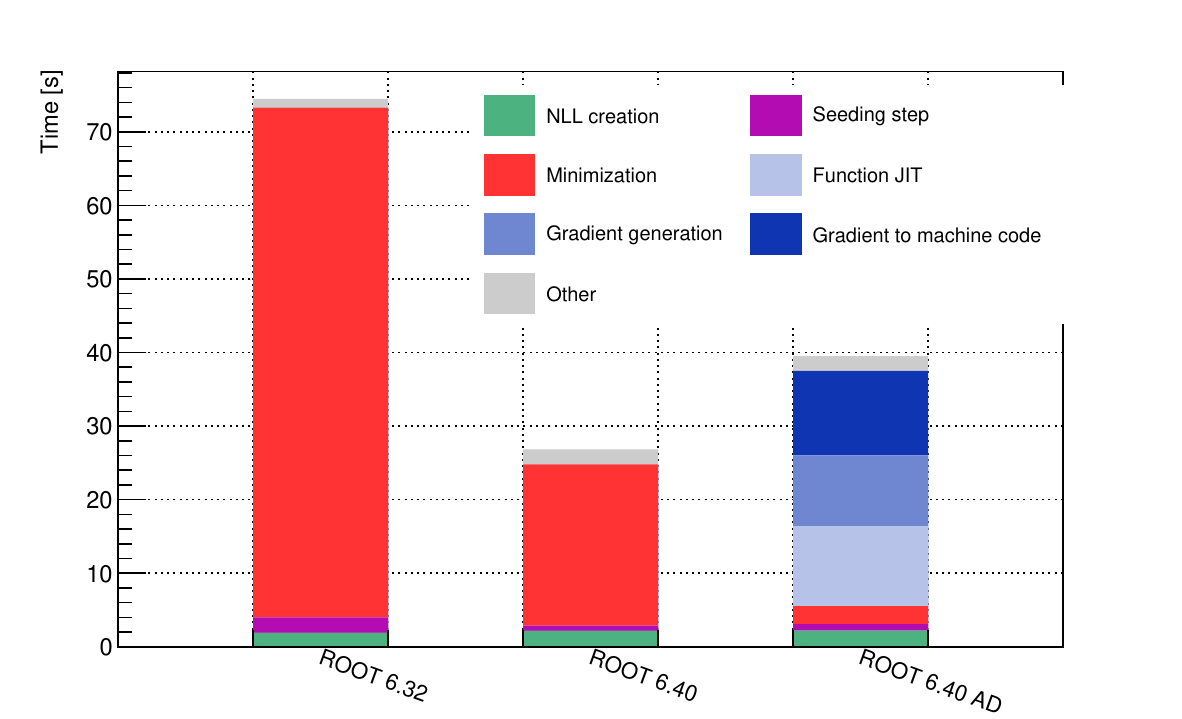}
\hfill
\includegraphics[width=0.48\textwidth]{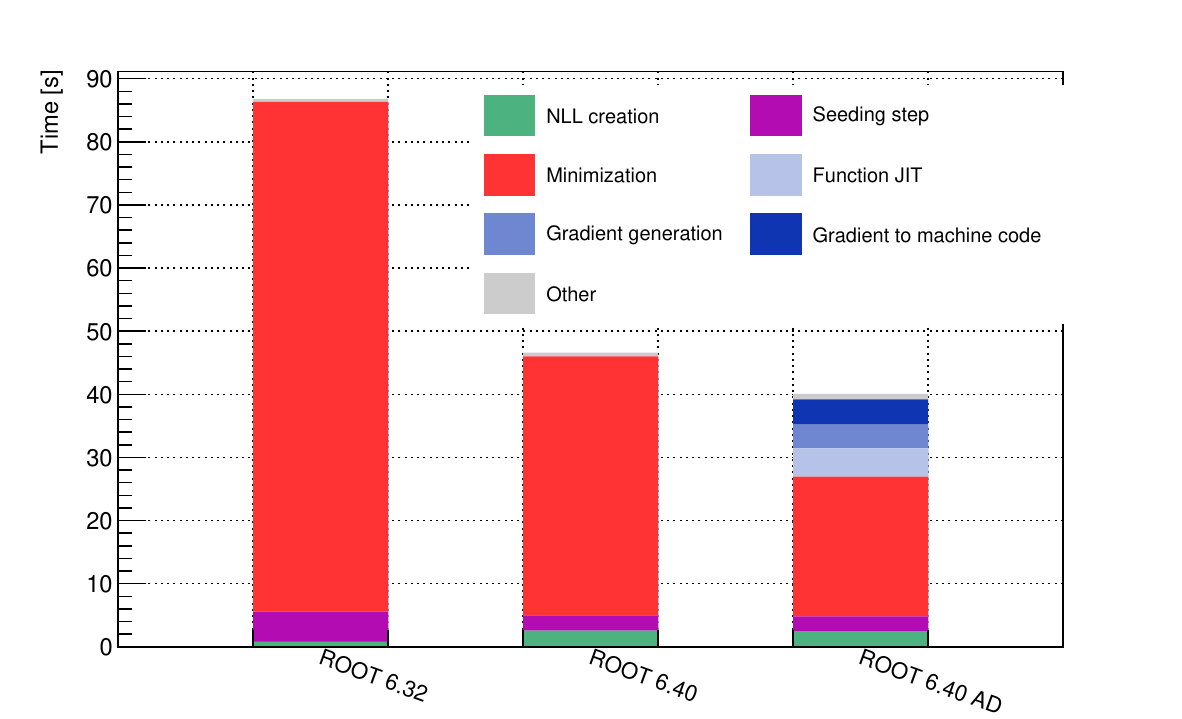}
\caption{Fit time composition for an ATLAS Higgs combination workspace (left) and
for the CMS Higgs boson observation likelihood (right), comparing ROOT 6.32,
ROOT 6.40 and ROOT 6.40 with automatic differentiation enabled.}
\label{fig:roofit_ad}
\end{figure}

Figure~\ref{fig:roofit_ad} illustrates the gains on two realistic benchmarks. On the
left, an ATLAS Higgs combination workspace with 49 channels and 739 parameters shows
a reduction of about a factor of ten in minimization time, so that the gradient is no
longer the bottleneck. On the right, the CMS Higgs boson observation
model~\cite{cms_collaboration_2024_c2948-e8875}, a more heterogeneous case with 672
parameters and 102 channels combining template histogram and analytical shape fits,
shows a visible but less pronounced improvement. In both cases the price of the
approach is the extra time spent generating the gradient, JIT-compiling the function
and translating it to machine code. That cost is paid only once, so the benefit is
largest in workflows that reuse the generated gradient many times, such as likelihood
scans and fits over toy datasets.

Several development directions follow. The RooFit team is working with CMS to
deploy AD within Combine, targeting codegen coverage for all classes used in binned
fits together with analytic minimization of Monte Carlo statistical nuisance
parameters inside the generated code. Analytic Hessian computation via Clad is on
the ROOT work plan for 2026, which would address the seeding-step bottleneck
visible in Figure~\ref{fig:roofit_ad}. Further work targets the reduction of JIT
overhead and the extension of codegen to additional RooFit primitives, including
the neural network components required by simulation-based inference workflows
(Section~\ref{sec:sbi}).

\section{Model preservation and interoperability: HS3}
\label{sec:hs3}

The statistical models used to derive the results of experimental analyses are of
considerable scientific value and are essential information for analysis
preservation and reuse~\cite{Cranmer:2021urp}. How they are built, however,
determines how readily they can be preserved.

Models in HEP fall into two broad categories~\cite{Cranmer:2021urp}. In the
\emph{open-world} approach, analysts define arbitrary custom components. This
offers maximum flexibility, but serialization relies on the \texttt{RooWorkspace}
container, and reading such a workspace requires access to the same custom
libraries used to create it, tying publication to a specific software stack. In the
\emph{closed-world} approach, the analyst is restricted to a finite, well-documented
set of building blocks and composition rules, which lends itself to a declarative
description decoupled from any particular implementation. Analyses based on
HistFactory~\cite{Cranmer:2012sba} are the canonical example: the \texttt{pyhf}
package introduced a JSON format encoding model and data in a single
ROOT-independent file, and ATLAS has published a growing number of full statistical
models on HEPData~\cite{Maguire:2017ypu} in this format. That format improves
portability and archival, but covers only the HistFactory model class.

The HEP Statistics Serialization Standard (HS3)~\cite{hs3} aims to cover both
worlds with a single JSON-based, framework-independent description of statistical
model, data and likelihood. HS3 is a standard rather than an implementation: it
specifies what a valid model description must contain and how each component maps
to a mathematical definition, leaving framework developers free to choose their own
realization. It also deliberately adopts statistical rather than HEP-specific
terminology, with the aim of making models accessible outside particle physics.

The format accommodates both a fully explicit description, in which every
distribution and function is spelled out, and compact declarative constructions that
encode a complete HistFactory channel in a few lines. Since the majority of LHC
analyses use HistFactory-class models, the compact form covers a large fraction of
use cases while remaining human-readable. Realistic models will nonetheless rarely be
written by hand, and HS3 therefore envisages a tooling ecosystem around programmatic
model generation and model-to-model transformations, such as combining likelihoods,
correlating nuisance parameters or reparametrizing for EFT interpretations.

Several implementations already target HS3 compliance. RooFit/RooStats covers close
to the full documented standard in C\texttt{++}. \texttt{pyHS3}~\cite{pyhs3}, a pure
Python implementation, is under active development. Support in CMS Combine is in
progress.

Two recently funded projects will drive adoption over the coming years. A Deutsche
Forschungsgemeinschaft project, \textit{Public Likelihood Combination}, supports a
proof-of-concept EFT interpretation of published HS3 models, with the explicit goal
of demonstrating end-to-end usability for theorists. The larger DEMOS initiative,
\textit{Democratizing Models}, of approximately 8 FTE over three years, targets
adoption beyond HEP through documentation, tutorials, community governance,
validation pipelines, converters, and a library of example models.

Challenges remain. An ongoing round-trip validation effort, exporting RooFit
workspaces to HS3 and re-importing them through \texttt{pyHS3}, has uncovered cases
where the ROOT exporter deviates from the evolving specification or omits
information needed for full model reconstruction. The specification itself still
contains ambiguities, including inconsistent naming conventions and unclear
semantics for generic distributions. Tooling support across the ecosystem is
incomplete: \texttt{pyhf} is actively adding HS3 import capabilities, and
preliminary work has demonstrated successful construction of JAX computation graphs
from \texttt{pyHS3}, but other tools do not yet support the standard. Broader
adoption also raises open questions, such as how to validate that an exported HS3
file faithfully reproduces the original model, and how to standardize systematic
uncertainty naming across experiments.

\section{Setup and roadmap in the experiments}
\label{sec:experiments}

\subsection{CMS: Combine}
\label{sec:cms}

Combine~\cite{CMS:2024onh} is the primary statistical inference tool in CMS, used
by approximately 92\% of analyses according to internal surveys. It takes an
end-to-end approach, from model building to final results, and provides a
command-line interface to RooFit and RooStats methods, supporting counting,
parametric (binned and unbinned) and template-based models. Analysts define models
through human-readable datacards specifying the primary and auxiliary likelihood
components; a physics model step then translates these into RooFit workspaces.
Combine implements the range of statistical procedures recommended by the CMS
Statistics Committee, including asymptotic limit setting~\cite{Cowan:2010js},
profile likelihood scans, goodness-of-fit tests and nuisance parameter impact
evaluation.

Combine also provides a number of custom components that optimize likelihood
evaluation beyond what standard RooFit offers. These add caching and constant-term
optimizations that avoid redundant evaluations, implement the Barlow--Beeston-lite
technique~\cite{Barlow:1993dm} so that the bin-by-bin Monte Carlo statistical
nuisance parameters of template fits are minimized analytically rather than passed to
the minimizer, and group processes so that systematic variations can be morphed
efficiently. Together they reduce the effective parameter count and improve both
memory usage and minimization time appreciably on large models. Combinations are
handled at the datacard level, before workspace creation.

CMS has established a systematic pipeline for publishing statistical models
alongside analysis results. Analysts store their datacards and ROOT files in a
central repository serving both analysis preservation and future combinations. Each
repository includes continuous integration tooling that validates datacards, checks
naming conventions, builds workspaces and runs standard statistical methods
automatically. Multiple review checkpoints ensure correctness before publication,
and the process runs in parallel with HEPData preparation so as not to delay papers.
Entries published on the CERN Document Server include the datacard, the ROOT
workspace, a table of systematic uncertainties, and the Combine version and commands
needed to reproduce the results.

As noted in Section~\ref{sec:hs3}, this mode of publication still relies on specific
software being available to reproduce the results. For this reason, current
development focuses, among other things, on the integration of the RooFit AD
codegen backend and on the implementation of HS3 support for model publication.

\begin{figure}[htbp]
\centering
\includegraphics[width=0.48\textwidth]{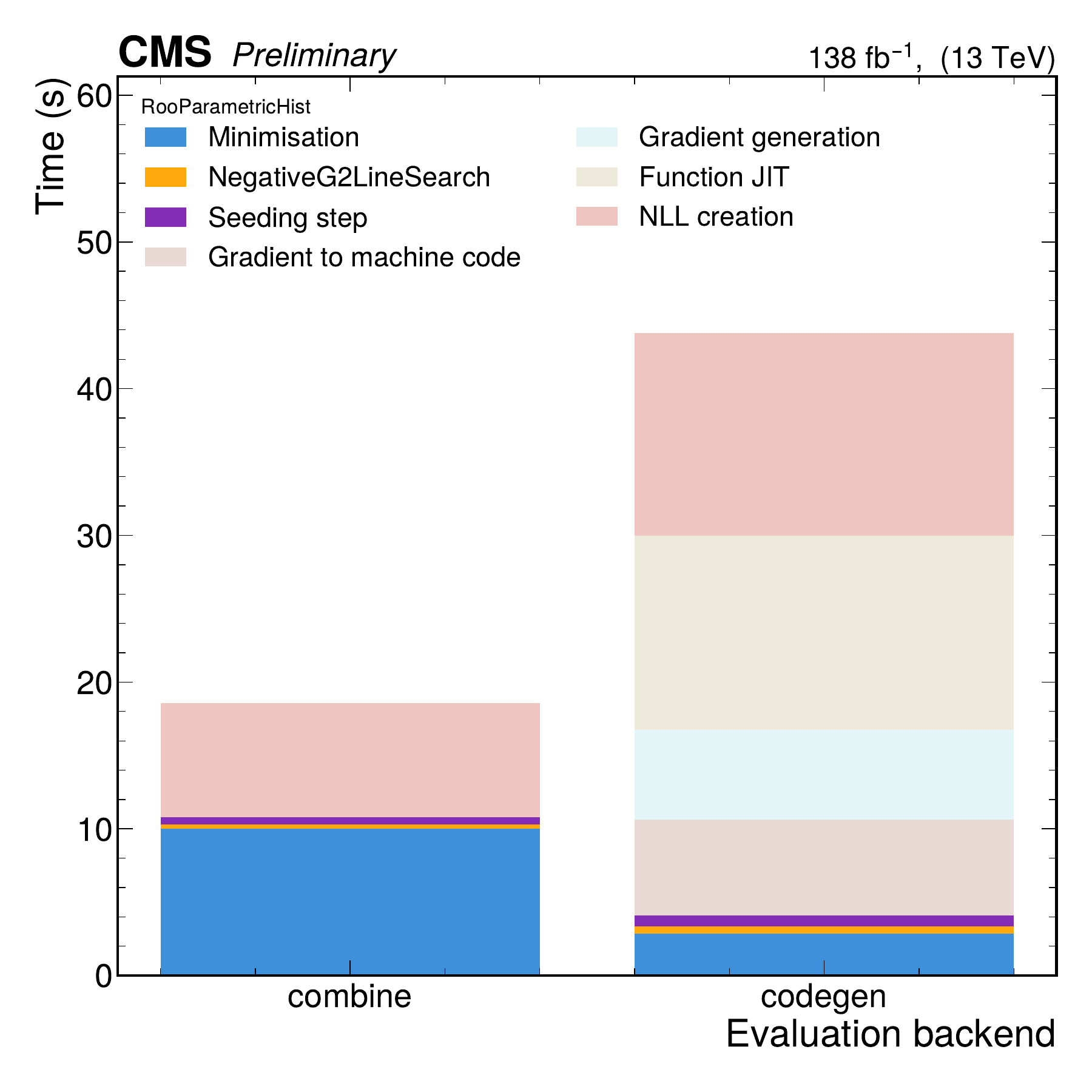}
\hfill
\includegraphics[width=0.48\textwidth]{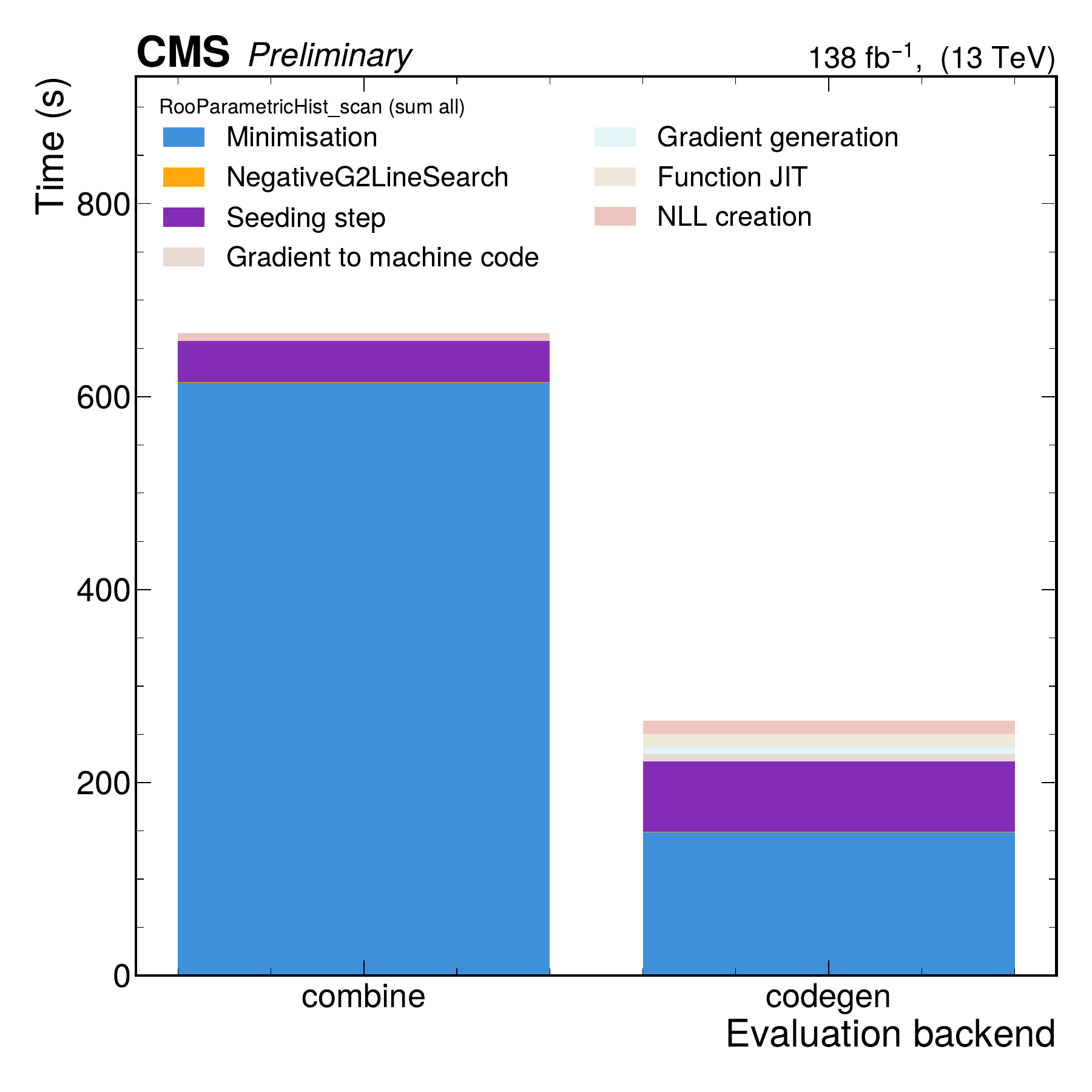}
\caption{Composition of the execution time for a CMS Combine benchmark, comparing
the default Combine evaluation backend with the codegen backend, for a single fit
(left) and for the sum of all the fits entering a parameter scan (right).}
\label{fig:combine_ad}
\end{figure}

The status of the AD integration in Combine is illustrated in
Figure~\ref{fig:combine_ad}, for a benchmark built from the statistical model of
Ref.~\cite{PRL131041801}. For a single fit, shown on the left, the total execution
time increases when the codegen backend is used: the minimization itself becomes
considerably cheaper, but that gain is outweighed by the one-off cost of creating the
negative log-likelihood (NLL), JIT-compiling the generated function, generating the
gradient and translating it to machine code. For a full parameter scan, shown on the
right, the same generated gradient is reused across many fits and the overall time is
reduced substantially. Codegen support within Combine is not yet optimized, and
further gains are expected as the integration proceeds.

\subsection{ATLAS: a diversified tooling landscape}
\label{sec:atlas}

ATLAS takes a more distributed approach, with of order ten tools coexisting. The
most widely used public ones are \texttt{TRExFitter} (C\texttt{++}/RooStats),
\texttt{HistFitter} (Python/RooStats) and \texttt{pyhf} (Python, ROOT-independent),
with \texttt{pyhf} adoption growing steadily; \texttt{XRooFit} and the
\texttt{quickFit}/\texttt{quickStats} family provide more general functionality.
Most of these tools act as wrappers around the HistFactory model, handling
preprocessing, assembling models from text or XML configuration files, and producing
diagnostic outputs such as pull and ranking plots, pre-fit and post-fit comparisons,
and likelihood scans.

The most common use case is the binned profile-likelihood fit. Unbinned likelihoods,
used in analyses such as $H\rightarrow\gamma\gamma$, top quark mass and B-physics
measurements, rely on dedicated RooFit-based frameworks, most of which are not
public. Combinations proceed through merging of HistFactory workspaces using
dedicated tools, while the BLUE method~\cite{Nisius:2020jmf} is often used for
non-HistFactory combinations.

ATLAS pioneered statistical model publication, releasing \texttt{pyhf} JSON
likelihoods since 2019. More recently, a number of analyses have published models
in HS3 format on HEPData.

Looking ahead, ATLAS identifies several priorities. Models will grow more complex at
the HL-LHC, with EFT fits requiring many simultaneous parameters of interest and
pseudo-experiments increasingly needed for limit setting; scalable tooling is
therefore essential, and the RooFit AD backend is already usable for non-custom
classes. Alternatives or improvements to Minuit and better Hessian computation rank
among the priorities, as does the adoption of HS3 as a common language across
frameworks in order to tackle larger combinations. The increasing role of ML-based
approaches, including simulation-based inference and unbinned unfolding, raises
further questions about result preservation and standardized formats.

\section{An emerging JAX-based ecosystem}
\label{sec:jax}

A growing amount of development is being directed towards JAX~\cite{jax2018github}. JAX is a NumPy-compatible array library that extends standard array programming with
composable functional transformations: just-in-time compilation (\texttt{jax.jit}),
automatic differentiation (\texttt{jax.grad}) and explicit vectorization
(\texttt{jax.vmap}). These transforms are applied at the level of a backend-agnostic
intermediate representation before translation to compiler targets such as XLA and
LLVM, which makes JAX naturally suited to CPU, GPU and TPU execution without code
changes. It integrates with the scientific Python ecosystem through the Array API
standard, connecting to NumPy, SciPy, Matplotlib, \texttt{uproot},
\texttt{boost-histogram} and \texttt{iminuit}. It also connects to a dedicated
machine learning and optimization ecosystem including
Equinox~\cite{kidger2021equinox}, Optax~\cite{deepmind2020jax} and
Optimistix~\cite{optimistix2024}. This combination makes it a natural fit for HEP
statistical workflows.

\begin{figure}[htbp]
\centering
\includegraphics[width=0.8\textwidth]{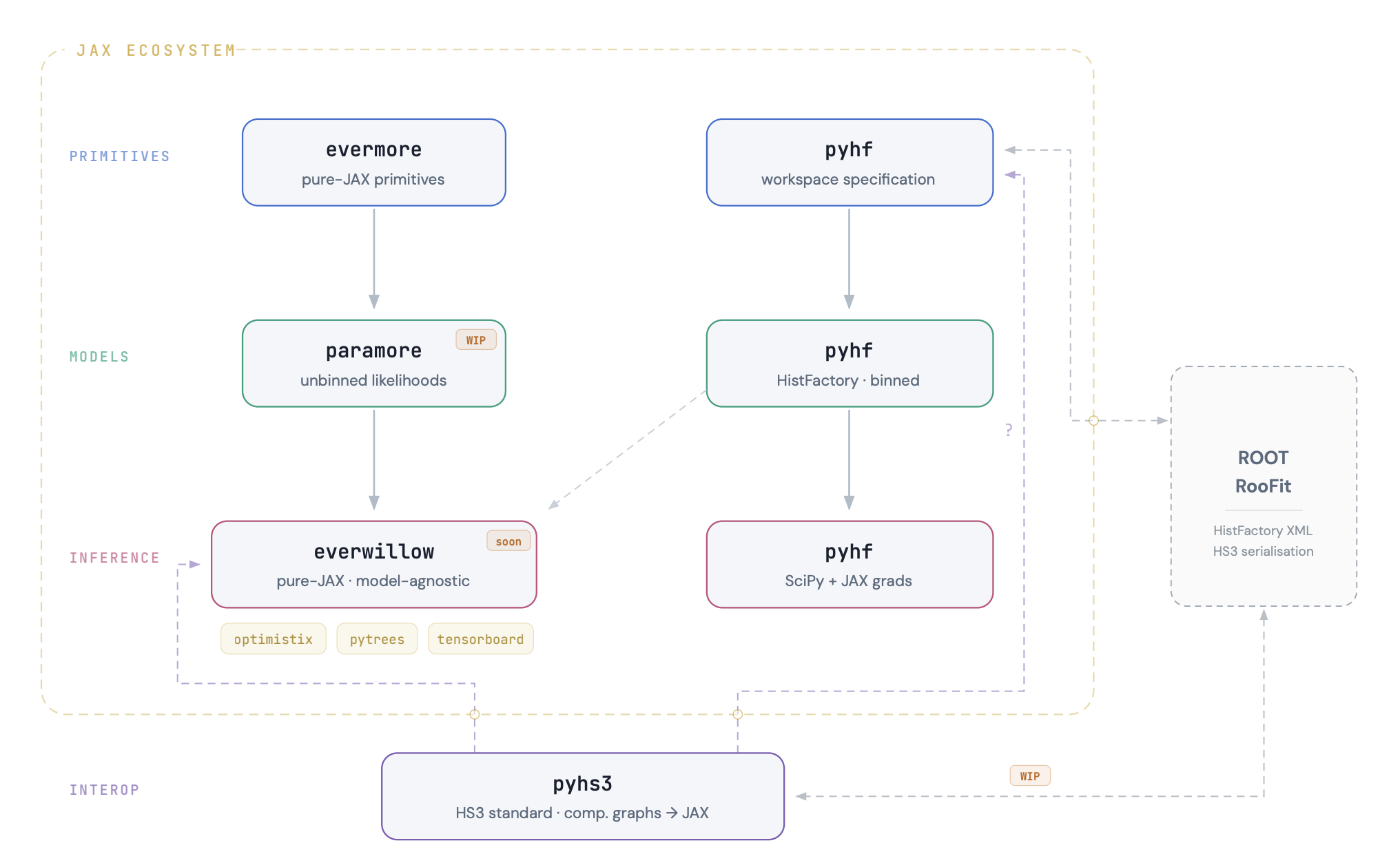}
\caption{Sketch of the JAX-based HEP statistical ecosystem under development.}
\label{fig:jax}
\end{figure}

Several HEP-specific libraries are being built on top of JAX, forming the layered
ecosystem sketched in Figure~\ref{fig:jax}, organized around model specification,
model building, inference and interoperability. At the specification level,
\texttt{pyhf}~\cite{pyhf_joss} implements HistFactory in pure Python with a JAX
backend and JSON-serialized workspaces, so that \texttt{jax.grad}, \texttt{jax.jit}
and \texttt{jax.vmap} apply directly to the likelihood.
\texttt{evermore}~\cite{evermore} provides lower-level JAX-native primitives for
building binned likelihoods, with parameters, modifiers and effects as first-class
objects built on PyTrees, the arbitrary nested Python containers that integrate
natively with JAX transforms; custom modifiers reduce to lambda functions, and
neural networks can be embedded directly as model components~\cite{flax2020github}.
At a higher level of abstraction, \texttt{paramore}~\cite{paramore}, a
work-in-progress package, extends \texttt{evermore} to unbinned likelihoods and
parametric models. \texttt{pyHS3}~\cite{pyhs3} provides the interoperability layer,
reading HS3 JSON workspaces and transpiling their computation graphs into
JAX-callable functions. The second version of \texttt{zfit}~\cite{Eschle:2019jmu},
currently being rewritten in JAX, is expected to fit naturally into the same
picture, providing tools for both binned and unbinned likelihoods.

For inference, \texttt{everwillow}~\cite{everwillow} is a likelihood-agnostic
minimization tool built entirely in JAX on top of Optimistix, supporting BFGS,
L-BFGS, MIGRAD, SIMPLEX and composable custom solvers. Because Optimistix optimizers
are themselves PyTrees, the entire minimization is end-to-end JIT-compilable, which
allows \texttt{jax.vmap} to be applied over full minimizations: hundreds of toy fits
or parameter scan points run in a single vectorized call. The tool requires only
that the likelihood be a JAX function with parameters expressed as PyTrees, and it
supports bounded parameter transforms and full fit introspection. Model combination
across frameworks is handled by the \texttt{statelib} utility, which uses PyTree
flattening to merge parameter structures from different sources into a single
unified state: each likelihood receives only its own parameter subtree, the
minimizer operates on the merged state, and the combined negative log-likelihood is
simply the sum of the component terms. Cross-model correlations between components
built with different frameworks follow directly from this merged-state representation.

The shared use of JAX PyTrees for representing parameters is what makes these
libraries naturally combinable and interoperable with one another. Extending the
same interoperability across the boundary with RooFit is a further goal, and here
the guiding principle is not to replace the established ecosystem but to integrate
with it. Two routes exist. At the \emph{model level}, an existing
RooFit model in C\texttt{++} can be serialized to HS3, which ROOT supports natively,
and transpiled into a JAX-callable function through \texttt{pyHS3}. At the
\emph{likelihood level}, negative log-likelihoods built in the two frameworks can be
summed and minimized within either world. A prototype demonstrating both directions exists\footnote{\url{https://gist.github.com/guitargeek/ea90e36abb46928ff0c2e976b7b69c85}}: in the first, a JAX-defined negative
log-likelihood is wrapped into a RooFit-compatible object through a small
C\texttt{++} bridge class, allowing it to be summed with a native RooFit NLL and
minimized with Minuit2, with analytic gradients flowing from both sides; in the
second, a RooFit NLL and its analytic gradient are exposed to JAX through callback
and custom differentiation mechanisms, so that the RooFit likelihood appears as a
standard differentiable JAX function that can be minimized with Optimistix. While
still at proof-of-concept stage on simple models, this establishes that combining
likelihoods across the two ecosystems is technically feasible and can preserve
end-to-end differentiability.


\section{Integration with simulation-based inference}
\label{sec:sbi}

The traditional statistical procedure in HEP builds the likelihood from histograms
or parametric models, which requires compressing high-dimensional data into
low-dimensional summary statistics, often at the cost of information loss.
Simulation-based inference (SBI) bypasses this limitation: when the likelihood is
intractable, hypothesis testing is performed using a likelihood ratio, or likelihood,
learnt from simulated samples with machine learning
techniques~\cite{Cranmer:2015bka,Brehmer:2018eca}. This allows analyses to work with
higher-dimensional observables and to capture correlations that a binned approach
would discard, leading in many cases to increased sensitivity. Analyses using SBI
techniques at various levels have already been published by
ATLAS~\cite{ATLAS:2025clx,ATLAS:2024jry} and CMS~\cite{CMS:2024ksn,CMS:2025dpp}.

From the point of view of the statistical tooling ecosystem, the principal challenge
is infrastructural: performing joint fits that combine SBI-based results with
traditional statistical workflows, which in practice means summing a simulated
negative log-likelihood with conventional NLL terms and minimizing the result.
Computational scalability and the serialization of SBI models are further open
issues. A concrete example of such a combination, bringing together an SBI analysis,
a semi-parametric approach and traditional analyses, has been carried out in the
ATLAS off-shell Higgs boson measurement~\cite{ATLAS:2024jry}; it was implemented in
both JAX and RooFit, in each case with \textit{ad hoc} solutions. Research and
development towards more scalable, out-of-the-box support is ongoing in both
ecosystems.

On the RooFit side, work is under way to allow interoperability between SBI and
traditional workflows. A new \texttt{RooONNXFunc}\footnote{\url{https://root.cern.ch/doc/v640/classRooONNXFunc.html}} class wraps any model expressed in
the ONNX open neural network exchange format into a RooFit object, and is fully
differentiable with Clad, so that learnt components can enter a standard RooFit
likelihood and participate in the AD pipeline described in Section~\ref{sec:ad}.

On the JAX side, a set of utilities for SBI analyses is being developed that
integrates with the wider JAX ecosystem, covering the pipeline from columnar data
access and event selection through the preparation of training inputs, the training
of calibrated density-ratio models, and model storage. The resulting simulated NLL
can then be combined with \texttt{evermore}-built likelihood terms and minimized
with \texttt{everwillow}, which places SBI components on the same footing as
conventional ones within the ecosystem described in Section~\ref{sec:jax}.


\section{Summary}
\label{sec:summary}

These are eventful times for statistical tooling in high energy physics. The growth
in data volume and analysis complexity expected at the HL-LHC will require
substantial improvements in the statistical inference step of HEP analyses, along
the four axes of performance, preservation, interoperability and integration with
machine learning.

The RooFit and RooStats stack remains the backbone of LHC statistical analyses, and
it is evolving quickly: automatic differentiation through the Clad-based codegen
pipeline already delivers substantial reductions in minimization time on realistic
benchmarks, with the caveat that the compilation overhead makes it most valuable in
workflows that reuse the generated gradient. The need for model
publication, serialization and reuse across frameworks has converged on HS3 as the
candidate standard, now backed by two funded projects and by implementations in both
the ROOT and Python worlds. The experiments are keeping pace: CMS has a production
publication pipeline and is integrating both AD and HS3 into Combine, while ATLAS
continues to publish models on HEPData and is moving towards HS3 as a common
language across its diversified tooling landscape.

In parallel, a new JAX-based ecosystem is growing, bringing complementary strengths
in differentiability, JIT compilation and hardware acceleration. Bidirectional
interoperability between RooFit and JAX has been demonstrated at prototype level,
showing that the two ecosystems can coexist and combine rather than compete.
Finally, the integration of simulation-based inference with standard analyses is an
active area in both worlds, with solutions being developed in RooFit and in the
Python-based systems alike.

\section*{Acknowledgments}

This work was supported by the U.S. National Science Foundation (NSF) under
Cooperative Agreement OAC-1836650 (IRIS-HEP).

\bibliography{proceedings}

\end{document}